\documentstyle[12pt,aasms4]{article}
\tighten

\newcommand\be{\begin{equation}}
\newcommand\ee{\end{equation}}

\input epsf
\lefthead{P. Kumar and S. Basu}
\righthead{Line asymmetry of solar p-modes}
\begin{document}

\title{\bf  Line asymmetry of solar p-modes:  \\
Properties of acoustic sources}

\author{Pawan Kumar and Sarbani Basu}
\affil{Institute for Advanced Study, Olden Lane, Princeton, NJ 08540, U. S. A.}

\begin{abstract}
The observed solar p-mode velocity power spectra are
compared with theoretically calculated power spectra 
over a range of mode degree and frequency.
The shape of the theoretical power spectra depends on the 
depth of acoustic sources responsible for the excitation
of p-modes, and also on the multipole nature of the
source. We vary the source depth to obtain the best fit
to the observed spectra. We find that quadrupole
acoustic sources provide a good fit to the observed
spectra provided that the sources are located between 700 km and 1050 km
below the top of the convection zone. The  dipole sources give a good fit
for  significantly shallower source, with a source-depth of  
between 120 km and 350 km.
The main uncertainty in the determination of
depth arises due to poor knowledge of 
nature of power leakages from modes with adjacent degrees,
and the  background in the observed spectra.
\end{abstract}


\section{Introduction}

The claim of Duvall et al. (1993) that solar p-mode line profiles
are asymmetric is now well established from the data 
produced by the Global Oscillation Network Group (GONG) and the
instruments aboard the Solar and Heliospheric Observatory (SoHO).
The unambiguous establishment of the assymetry is
imporatnt since fitting a symmetric profile to the assymetric line
in the power-spectra can give errorneous results for the 
solar eigenfrequencies (cf. Nigam \& Kosovichev 1998).
In the last few years a number of different aspects of
line asymmetry problem have been explored and there 
appears to be a general consensus that the degree
of line asymmetry depends on the depth of acoustic sources
responsible for exciting solar p-modes (cf. Duvall 
et al. 1993; Gabriel 1995; Abrams \& Kumar 1996; Rosenthal 1998). 
We make use of the best available observed power spectra, for low 
frequency p-modes, and solar model,
to determine the location and nature of sources. 
One of the main differences between this work and others is that
we determine the source-depth using a realistic solar model
and taking into account the fact that the observational profiles
of a given degree have a substantial contribution of power from 
modes of neighbouring degrees. The effect of
$\ell$-leakage on the line asymmetry and on the background of the
power spectra which affects the determination of source depth is considered
in \S2. The main results are summarized in \S3.

\section{Theoretical calculation of line asymmetry}

The calculation of power spectra is carried out using the
method described in Abrams \& Kumar (1996) and Kumar (1994). 
Briefly, we solve the coupled set of linearized mass, 
momentum and entropy equations, with a source term,
using the Green's function method. We parameterize the source by
two numbers -- the depth where the source peaks and the radial
extent (the radial profile is taken to be a Gaussian function),
instead of taking the source as given by the mixing length theory
of turbulent convection. Power
spectra for different multipole sources are calculated
using the following equation

\be
P(\omega) = \left| \int dr S(r,\omega)\; {d^n G_\omega \over dr^n}\right |^2,
\ee
where $n=0$ for dipole and $1$ for quadrupole sources, and $G_\omega$ 
is the Green's function for the linearized set of non-adiabatic wave equations.
Physically, dipole sources produce acoustic waves by applying a time
dependent force on the fluid whereas only fluctuating internal stresses
are associated with quadrupole sources, and there is no associated
net momentum flux.

The power spectrum calculated using equation (1) is asymmetric, and the 
amount of asymmetry is a function of the source depth (cf. Abrams \& 
Kumar 1996). Thus by matching the asymmetry of the theoretical spectrum 
to that of the observed, we can estimate the depth of the sources that
excite solar p-modes. 

Other than the source depth, for any given mode  there are three free 
parameters in the fit --- the amplitude, the line width and the background. 
We normalize the observed and model amplitudes to unity. The background 
is assumed to be frequency independent over the narrow range of 
frequencies over which we carry out the fit.

The solar model used in this work  is a standard model of the present Sun.
It was constructed with OPAL opacities (Iglesias \& Rogers 1996) supplemented by
low temperature opacities of Kurucz(1991), and the OPAL
equation of state (Rogers, Swenson \& Iglesias 1996) was used. Convective flux
is calculated using the formulation of Canuto \& Mazzitelli (1991), and
the photospheric structure is calculated using the empirical $T-\tau$ 
relation of Vernazza et al.~(1981).  To check model dependencies
we have also used a similar model which uses the mixing length formalism
to calculate convective flux. We have also used Model S of
Christensen-Dalsgaard et al. (1996) and the old Christensen-Dalsgaard model
used in Kumar (1994, 1997) which had earlier been used to determine the
source depths.

The observed power spectra used in this work are the 144 day data from
the  Michelson Doppler Imager (MDI) of the Solar Oscillation Investigation
(SOI) on board SoHO and the data obtained by GONG during months 4 to 10
of its operation.

\subsection{The source depth without $\ell$-leakage}

The calculated power spectrum superposed on the observed 
spectrum is shown in Fig.~1. The calculated spectrum in fig. 1
does not include power leak from modes of adjacent degrees.

We define the goodness-of-fit by the merit function (cf. Anderson, 
Duvall \& Jeffries 1990)
\be
F_m={1\over N} \sum_{i=1}^N\left({O_i-M_i}\over M_i\right)^2
\ee
where,  the summation is over all data points, $O_i$ is the observed power 
and $M_i$ the model power. The quadrupole and dipole sources give 
very similar line profiles. The figure  of merit are also very 
similar 0.0069 for quadrupole and 0.0073 for dipole sources for the 
fit to the $\ell=35$ mode, in the frequenciy range $\pm 6\; \mu$Hz
from the peak. For the $\ell=60$ mode, $F_m$ is 0.0053 for the quadrupole
source and 0.0047 for the dipole.

However, the source depths obtained for the two types of sources are 
very different. For the same source depth dipole and quadrupole sources 
give rise to different sense of line asymmetry. 
The source depth for the   spectrum shown in fig. 1 is about
1050 km for the quadrupole source and the 350 km for the dipole source.
The depth required seems independent of the degree of the
mode in the range where we attempted the fit ($\ell$=35-80).
Though there appears to be a small dependence on the 
frequency of the mode, with higher frequency modes $\sim$ 3 mHz
requiring slightly shallower sources. The difference is within the
uncertainty.

Note that there is some 
difference between the observed and the
theoretical spectra in the wings of the lines.  We find that
we cannot fit the line wings properly without taking $\ell$-leakage
into account.

\subsection{The source depth with $\ell$-leakage}

The observed spectra for modes of a desired degree
contains leakage from neighboring 
$\ell$-modes as a result of partial observation of the
solar surface. 

A rough  estimate for $\ell$-leakage can be 
obtained from the amplitudes of different $\ell$-peaks
at low frequencies. Inclusion of $\ell$-leakage in the
theoretical calculations decreases asymmetry of the
the model. This can be seen in Fig.~2.
Moreover, it can be seen that $\ell$-leakage
contributes to the background of the spectra. This
implies that the observed spectrum can be fitted with a theoretically
calculated power-spectrum with a shallower source-depth
and a significantly smaller background term than that was 
used in \S 2.1.  As a result,  source-depth determined in the previous 
section is just an upper limit to the depth. 

In Fig.~3 we show the fit to the observed spectrum
of an $\ell=35$ mode when
$\ell$-leakage is taken into account while calculating the 
theoretical spectrum. The source depths needed are
950 km for a quadrupole source and 300 km for a dipole
source. We have taken the leakage from mode of degree ($\ell-2$) to be
10\% of its power, the leak from $\ell-1$ mode is 45\% of its
power, the $\ell+1$ mode leaks 75\% of its power and 
the $\ell+2$ mode leaks 25\% of its power into the 
power spectrum of the mode under consideration. These
numbers were estimated from the observed power spectrum
for the mode obtained by the GONG network.
The best fit theoretical curves, for both dipole and quadrupole  
sources, provide a much better match to the observed spectrum than
the case where the leakage was ignored.  The figures of merit are
0.0028 for the quadrupole source and 0.0025 for the dipole source.
So unfortunately it is still not possible to say whether the
excitation sources in the Sun are quadrupole or dipole based
on the observed asymmetry of low frequency p-modes.

To get a lower limit to the depth, we considered an extreme
case where modes of $\delta\ell=\pm1$ and $\pm2$ leak all their 
power into the mode under consideration, and find that for
quadrupole sources we require a source depth of 700 km, and
for a dipole source, a source depth of 120 km is needed.

\section{Discussion}
  
The calculation of power spectra using the method described in
Kumar (1994) has been shown previously to fit the high frequency
velocity power spectra (cf. Kumar 1997). In this paper we have shown
that the same method can be used to calculate the power spectrum with
observed line shapes for low and intermediate frequency modes also.
The variable parameter is the source depth. The depth required to fit
the observed low frequency p-mode spectra depends mainly on the nature
of the sources; quadrupole sources have to be very deep --- between 700
and 1050 km while dipole sources need to be  relatively shallow --- between
120 and 350 km. The main uncertainty in the source depths arises due to
the lack of accurate knowledge of power leakages into the spectrum
from modes of adjacent degrees. We find that using the low
frequency data it is not possible to say whether the sources that
excite solar oscillations are dipole in nature or are quadrupolar.
The source depth can have some
latitude dependence. However, the observed spectra are m-averaged
to improve the signal which precludes the determination of possible
latitude dependence.

Acoustic waves of frequency 2.2mHz are evanescent at depths less than 
approximately 900 km. So it appears, according to our best fit model,  
that the source of excitation for low frequency waves lies in the 
evanescent region of the convection zone.

The frequencies of peaks in the theoretically calculated power spectra
are shifted with respect to the non-adiabatic eigenfrequencies of
corresponding p-modes by approximately 0.1$\mu$Hz for modes of 2 mHz,
and 0.2$\mu$Hz at 3 mHz ($35\le\ell\le80$).

We have repeated some of the calculations with a model constructed
with the conventional mixing length theory, and find that the
source depth  decreases by about 50 Km, which is much
smaller than the other uncertainties.  We find
the same using Model S of Christensen-Dalsgaard et al. (1996). Since
Model S too is constructed with using the mixing length theory, the
difference in source depth must be a results of the difference in
surface structure because of the two different convection formalisms.
It is known that  models constructed with the
Canuto-Mazzitelli formulation of convection have frequencies that are
closer to solar frequencies than models  constructed with the
standard mixing length formalism (Basu \& Antia 1994).

The fit to the high frequency part of the observed spectra, for
peaks lying above the acoustic cutoff frequency of $\sim 5.5$ mHz, 
is provided by sources lying at somewhat shallower depth, although the
GONG/MDI data we have used does not show peaks beyond $\sim 7.5$mHz
and the signal to noise is not very high which precludes us from
assigning a high significance to this result. Kumar (1994 \& 1997)
using  South pole spectra which had clear peaks extending to
10 mHz, had found that quadrupole sources lying about 140 km below
the photosphere provide a good fit to the entire high frequency 
power spectra, but Kumar used an older solar model. With the
model used by Kumar (1994 \& 1997), we find that the observed
line profiles of low frequency p-modes are well modelled when
quadrupole sources are placed at a very shallow depth of order
of 200 Km. Thus we conclude that the source depth determination
is quite sensitive to the inaccuracies of the solar model near the
surface. Since the newer models are much better than the old ones,
perhaps the source depth determination using the newer models
has less systematic error.

In this paper we have concentrated on the velocity power spectra
of solar oscillations. In a companion paper, we consider the 
question of reversal of asymmetry in the intensity power
spectrum relative to the  velocity power spectrum.

\acknowledgments

This work  utilizes data from the Solar Oscillations
Investigation / Michelson Doppler Imager (SOI/MDI) on the Solar
and Heliospheric Observatory (SoHO).  SoHO is a project of
international cooperation between ESA and NASA.
This work also utilizes data obtained by the Global Oscillation
Network Group (GONG) project, managed by the National Solar Observatory, a
Division of the National Optical Astronomy Observatories, which is
operated by AURA, Inc. under a cooperative agreement with the
National Science Foundation. The data were acquired by instruments
operated by the Big Bear Solar Observatory, High Altitude Observatory,
Learmonth Solar Observatory, Udaipur Solar Observatory, Instituto de
Astrofisico de Canarias, and Cerro Tololo Inter-American Observatory. 

We thank John Bahcall for his comments and suggestions. We thank
J{\o}rgen Christensen-Dalsgaard for providing the Model S variables
needed for this work. PK is partially supported by  NASA grant 
NAG5-7395, and SB is partially supported by an AMIAS fellowship.

\begin{figure}
\plotone{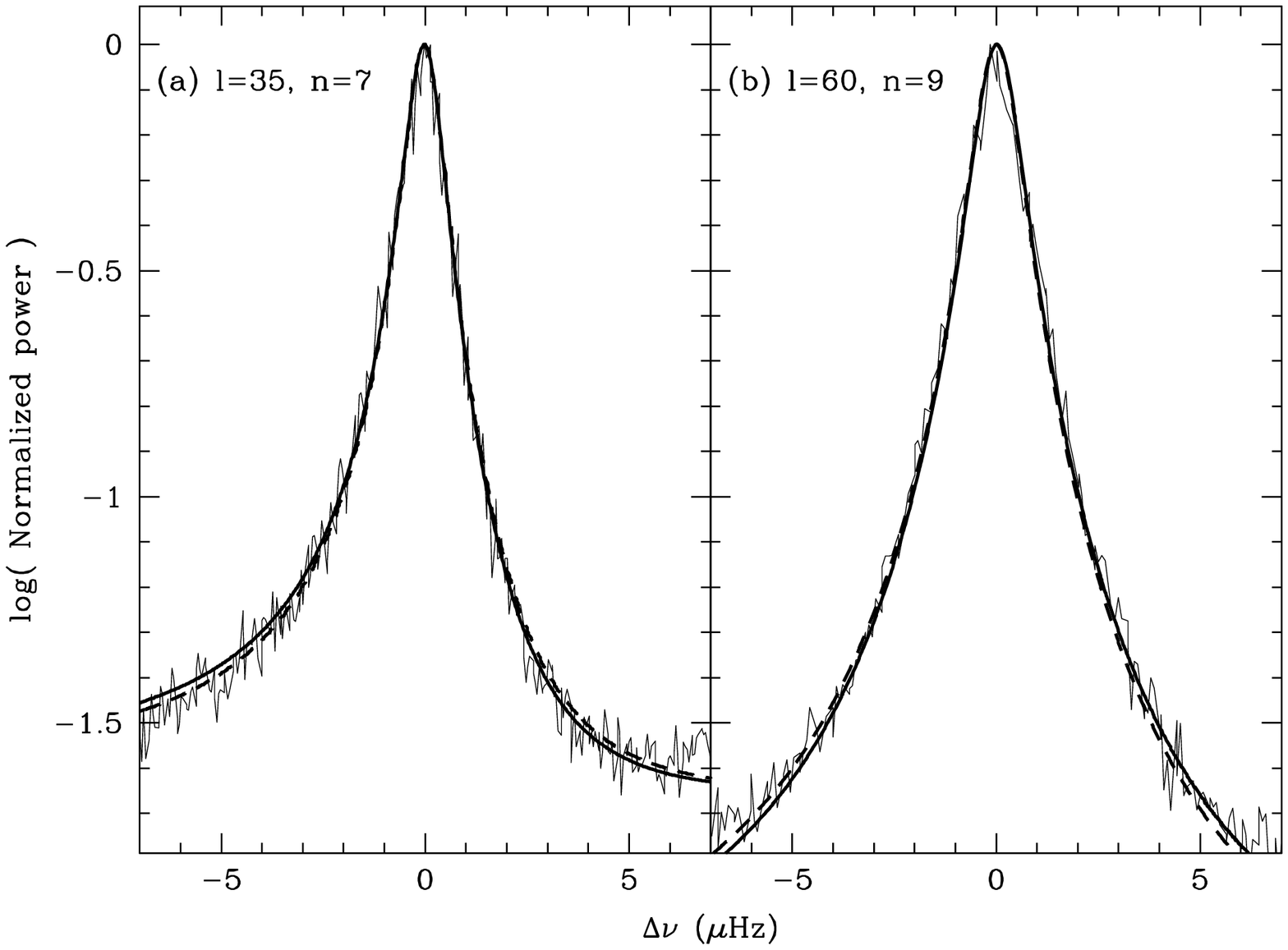}
\figcaption{ The calculated line profile superposed on observed line
profiles. The observed line profiles are shown by the thin 
continuous lines. They have been obtained by averaging the spectra of
all azimuthal orders, $m$, of the given mode after frequency shifts to account
for rotational splitting. The theoretical profiles obtained with
a quadrupole source is shown as the heavy continuous line and 
that for the dipole source is shown by the heavy dashed line.
The quadrupole source is at a depth of 1050 km and the
dipole at depth of 350 km from the top of the convection
zone.
The data for the $\ell=35$ mode are GONG data, while the 
$\ell=60$ data are from MDI.
}
\end{figure}

\begin{figure}
\plotone{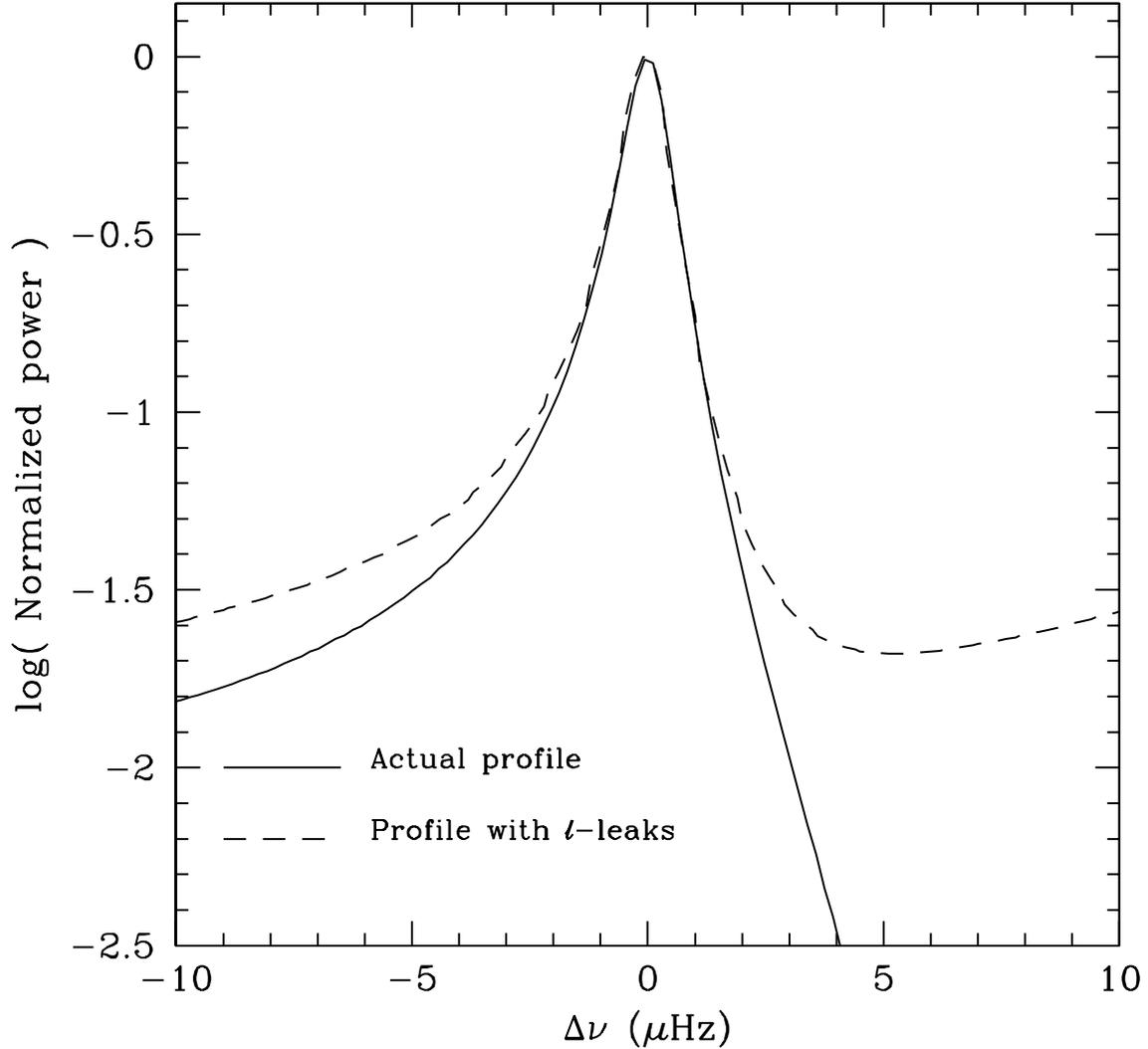}
\figcaption{The line profile calculated without leaks (continuous line)
compared with that obtained when power leakage from $\ell=\pm1$ and $\ell=\pm2$
modes is included (dashed line). The source depth is the same for both cases.
 Note that
the profile with leakage is more symmetric, thereby mimicking the line-profile 
obtained a deeper source when a  power background is assumed.
}
\end{figure}

\begin{figure}
\plotone{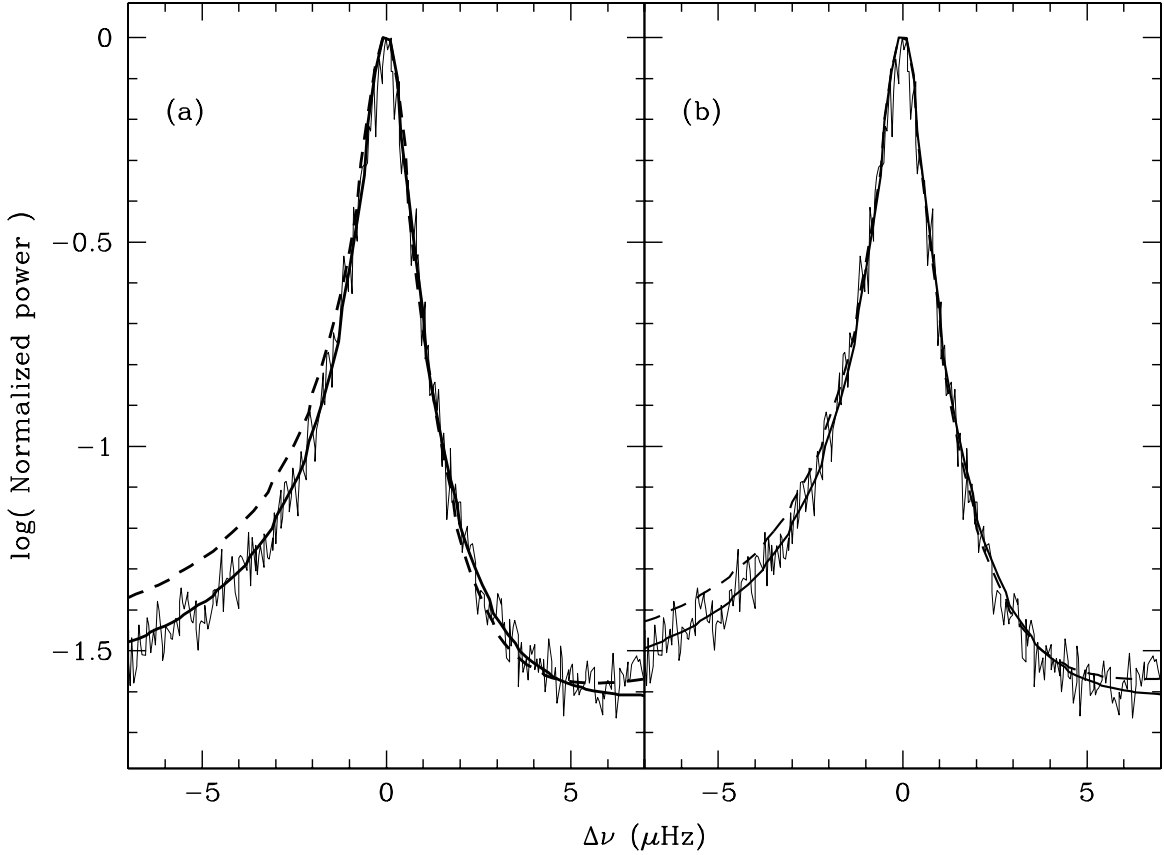}
\figcaption{Line profiles calculated with $\ell$-leakage taken into account 
superposed on the observed line profile of an $\ell=35$, $n=7$ mode that
has a frequency of about 2.22 mHz.
Panel (a) shows the results with quadrupole sources. The continuous line
is a source at depth 950 while the dashed line is a source at
depth 500km. Both have a small, frequency-independent background
added. The background is of the order of 1\% of the peak power.
Note that it is not possible to fit the observed data
with a shallow source by simply changing the background since
the line shape does not match the observed profile..
Panel (b) are results with dipole sources. The continuous line
is for a source at a depth of 300 km and the dashed line is
for a source depth of 120 km below the top of the convective
zone.
}
\end{figure}

\end{document}